\documentclass[conference]{IEEEtran}
\IEEEoverridecommandlockouts

\usepackage{xcolor}
\usepackage{multirow}
\usepackage{subfigure}
\usepackage{bbding}
\usepackage{makecell}
\usepackage{mathtools}
\usepackage[super]{nth}
\usepackage{url}
\usepackage{pifont}

\usepackage{algorithm}
\usepackage{algpseudocode}
\algtext*{EndFor}
\algtext*{EndFunction}
\algtext*{Until}
\algtext*{EndIf}
\algtext*{EndWhile}
\usepackage{amsmath}
\usepackage{array}
\usepackage{tablefootnote}

\usepackage{soul}
\usepackage{siunitx}
\usepackage{multicol}

\def\BibTeX{{\rm B\kern-.05em{\sc i\kern-.025em b}\kern-.08em
    T\kern-.1667em\lower.7ex\hbox{E}\kern-.125emX}}

\newcommand{\yao}[1]{\textcolor{black}{#1}}

\newcommand{\mm}[1]{\textcolor{black}{#1}}
\newcommand{\mmi}[1]{\textcolor{black}{#1}}

\begin{document}

\pdfpagewidth=8.5in
\pdfpageheight=11in

\newcommand{\iscasubmissionnumber}{207}

\pagenumbering{arabic}

\title{\huge{ICP: Exploiting \underline{I}nstruction \underline{C}orrelation for \underline{P}refetching Irregular Memory Accesses}}

\DeclareRobustCommand*{\IEEEauthorrefmark}[1]{%
\raisebox{0pt}[0pt][0pt]{\textsuperscript{\footnotesize\ensuremath{#1}}}}

\author{\IEEEauthorblockN{Mengming Li\IEEEauthorrefmark{1},
Chenlu Miao\IEEEauthorrefmark{2}, Buqing Xu\IEEEauthorrefmark{1}, Qijun Zhang\IEEEauthorrefmark{1}, Xiangfeng Sun\IEEEauthorrefmark{1}, Ceyu Xu\IEEEauthorrefmark{1}, Yuan Xie\IEEEauthorrefmark{1} \\ Wenkai Li\IEEEauthorrefmark{1}, Shang Liu\IEEEauthorrefmark{1} and
Zhiyao Xie\IEEEauthorrefmark{1*}\thanks{*Corresponding Author}}
\vspace{.05in}
\IEEEauthorblockA{\IEEEauthorrefmark{1}The Hong Kong University of Science and Technology, \IEEEauthorrefmark{2}Independent Researcher \\
mengming.li@connect.ust.hk, chenlu.miao@outlook.com, \{bxuax, qzhangcs, xsunbv\}@connect.ust.hk, \\ 
\{eeentropy, yuanxie\}@ust.hk, \{wlidm, sliudx\}@connect.ust.hk, eezhiyao@ust.hk\\}\vspace*{-.99cm}}

\maketitle
\thispagestyle{plain}
\pagestyle{plain}

\begin{abstract}

Irregular memory accesses pose challenges for effective and efficient data prefetching. While temporal prefetchers have recently shown promise for irregular memory access patterns, their effectiveness fundamentally depends on temporal address recurrence and large metadata storage. When memory addresses exhibit weak or no recurrence, as in indirect memory accesses, temporal prefetchers achieve limited performance gains while incurring substantial storage overhead.

This paper proposes Instruction-Correlation Prefetching (ICP), a new hardware prefetching mechanism that exploits instruction-level correlations rather than memory-address correlations to handle irregular memory accesses. ICP observes that although memory addresses may not repeat, the instructions generating them often recur with stable data-dependency relationships. By learning these persistent instruction correlations, ICP speculatively computes and prefetches future irregular accesses using the execution results of their correlated predecessors. Across irregular SPEC CPU and GAP benchmarks, ICP outperforms the state-of-the-art temporal prefetcher Triangel by 14.0\% and the indirect prefetcher DMP by 6.0\%, while requiring only \mmi{2.1 KB} of hardware storage, over three orders of magnitude smaller than temporal prefetchers.

\end{abstract}

\section{Introduction}
\label{sec:intro}

Cache prefetching, a long-standing technique for mitigating the “memory wall” problem~\cite{wulf1995hitting}, has been extensively explored as a means to improve processor performance. The fundamental challenge lies in efficiently predicting the diverse and complex memory access patterns that occur across different applications. To address this challenge, researchers have proposed various prefetching techniques tailored to distinct classes of access patterns~\cite{alecto,gerogiannis2023micro,alcorta2023lightweight,zhang2022resemble,pakalapati2020bouquet,kondguli2018division}.
These prefetchers include stream prefetchers~\cite{jouppi1990improving,he2022dsdp,hur2006memory}, stride prefetchers~\cite{baer1991effective,dahlgren1995effectiveness,kim1997stride}, spatial prefetchers~\cite{somogyi2006spatial,kim2016path,michaud2016best,navarro2022berti,bakhshalipour2019bingo,bera2019dspatch,pugsley2014sandbox,shevgoor2015efficiently,ishii2009access,jiang2022merging}, indirect prefetchers~\cite{zhang2024rpg2,jamilan2022apt,ainsworth2017software,khan2021dmon,callahan1991software,chen1991data,gornish1990compiler}, and temporal prefetchers~\cite{jain2013linearizing,nesbit2004data,wu2019efficient,wenisch2009practical,wu2019temporal,bakhshalipour2018domino}.

Among various prefetchers, \textbf{temporal prefetchers} represent a promising technique for addressing \emph{irregular memory accesses}, where conventional prefetchers (e.g., stream, stride, spatial prefetchers) fall short. The core idea behind temporal prefetchers is to record previously accessed memory addresses and their correlated successor addresses as \emph{metadata}. When a recorded address reappears during execution, prefetchers use this metadata to prefetch its correlated successor addresses.

Practically, the effectiveness of temporal prefetchers depends on two factors: 1) whether memory addresses involved in irregular access patterns exhibit temporal recurrence; 2) if such recurrence exists, how many of those addresses can be retained in table until their next appearance. These two factors fundamentally constrain the capability of temporal prefetchers.

\textbf{Factor 1. Not all irregular memory accesses exhibit temporal recurrence.} Currently, two classes of irregular memory accesses are commonly found in programs: \emph{pointer accesses} and \emph{indirect memory accesses}. Our evaluation shows that while temporal prefetchers are generally effective for pointer-based accesses, they fail to capture most indirect memory accesses\footnote{Detailed experimental results are presented in Section~\ref{sec:eval}.}. Specifically, we evaluate a state-of-the-art temporal prefetcher on irregular sparse workloads dominated by indirect memory accesses~\cite{fu2024differential, fu2025magellan}. Our results show that it achieves only a 5.9\% performance improvement over the baseline, whereas an indirect prefetcher yields a substantially higher 42.9\% improvement. Furthermore, we observe that the metadata stored by temporal prefetchers exhibit a low reuse ratio ($10^5$ less than our solution) when processing indirect memory accesses. These observations all highlight their inherent limitation when memory addresses show weak or no temporal recurrence.


\begin{figure}[t]
\centering
\includegraphics[width=.99\linewidth]{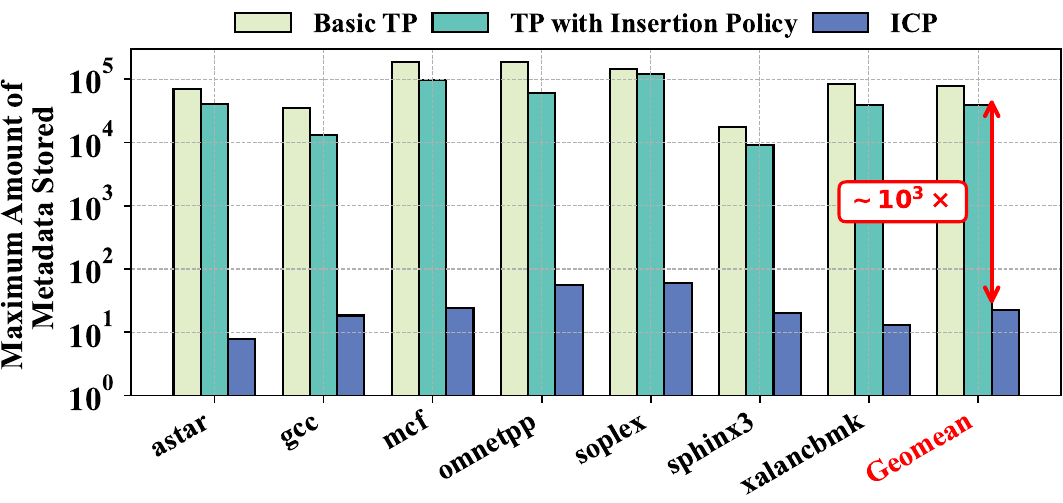}
\vspace{-.25in}
\caption{Comparison of the maximum amount of metadata stored by ICP and temporal prefetchers (TP). ICP reduces the metadata overhead by three orders of magnitude.}
\vspace{-.2in}
\label{fig:figure1}
\end{figure}

\textbf{Factor 2. Temporal prefetchers suffer from significant storage overhead.} Although temporal prefetchers effectively handle certain irregular memory access patterns (e.g., pointer-based accesses), they suffer from substantial metadata storage requirements, which are inherently difficult to reduce. This limitation arises because temporal prefetchers must record a large number of memory address correlations to maintain sufficient coverage of recurring access patterns. To illustrate this challenge, we evaluate the maximum metadata storage required by temporal prefetchers on irregular SPEC CPU benchmarks. Two representative classes of temporal prefetchers are considered: \emph{basic temporal prefetchers} (e.g., Triage~\cite{wu2019temporal}) and \emph{enhanced temporal prefetchers} that incorporate non-recurrence metadata filtering strategies (e.g., Triangel~\cite{ainsworth2024triangel}). As shown in Figure~\ref{fig:figure1}, both designs require metadata storage of the same order of magnitude, demonstrating that the large storage footprint is a fundamental limitation of temporal prefetching rather than an artifact of a specific design.

In response to the fundamental limitations of temporal prefetchers, we propose \textbf{ICP}, a new prefetching technique that exploits general instruction-level correlations rather than memory address correlations to handle irregular memory accesses. The core insight of ICP is that, although many irregular memory accesses lack temporal recurrence at the address level, the instructions generating these accesses often recur with consistent patterns during program execution. For example, consider two dependent load instructions, $\textit{PC}_\text{i}\!:\!ld\ a1,0(a0)\rightarrow \textit{PC}_\text{j}\!:\!ld\ a2,0(a1)$. Here, $\textit{PC}_\text{i}$ issues irregular memory accesses. While the specific memory addresses accessed by $\textit{PC}_\text{i}$ and $\textit{PC}_\text{j}$ may never repeat, their data-dependency relationship $a1$ remains stable across dynamic instances. By learning such persistent instruction correlations, ICP can speculatively compute and prefetch the memory addresses of irregular instructions using the execution results of their correlated predecessors.

ICP not only handles irregular memory accesses where temporal prefetchers fall short but also significantly reduces metadata storage requirements. As shown in Figure~\ref{fig:figure1}, ICP’s metadata footprint is over three orders of magnitude smaller, i.e., $<\!0.1\%$ of that required by temporal prefetchers. This efficiency stems from the fact that a single memory-access instruction may generate a vast number of dynamic requests. Consequently, one instruction-level correlation in ICP implicitly captures thousands or even millions of memory address correlations. By storing correlations at the instruction level rather than the address level, ICP achieves dramatically lower storage demands while maintaining comprehensive coverage of memory access patterns.

ICP is a \emph{hardware-based} prefetching mechanism for irregular memory accesses. It consists of two primary components: PC Correlation Detection and Prefetching with PC Correlations. The PC Correlation Detection stage performs two key tasks:
(1) it identifies memory instructions that are likely to generate irregular memory accesses (denoted as $\textit{PC}_\text{suc}$) and their corresponding predecessor instructions that produce the required input values (denoted as $\textit{PC}_\text{pre}$); and
(2) it constructs the data-dependency paths starting from each $\textit{PC}_\text{pre}$ and ending at its dependent $\textit{PC}_\text{suc}$: ($\textit{PC}_\text{pre}$, $\textit{PC}_\text{suc}$), and records them in the table. The Prefetching with PC Correlations stage then leverages cache line responses associated with $\textit{PC}_\text{pre}$ to speculatively compute memory addresses and prefetch data for the corresponding $\textit{PC}_\text{suc}$. To enhance both prefetch coverage and timeliness, ICP simultaneously exploits cache lines fetched by demand and prefetch requests to drive the process.

\mmi{\textbf{Improvement over other prefetching solutions exploiting speculative instruction execution.} ICP improves upon prior works by providing stronger generality for irregular memory accesses while maintaining moderate hardware complexity. This improvement stems from two key aspects of its design: (i) how it defines the speculated instruction chains and (ii) how it performs speculative execution to generate prefetches.}

\mm{\textbf{Definition of speculated instruction chains.} State-of-the-art \emph{indirect prefetchers} (e.g., Tyche~\cite{xue2024tyche} and DMP~\cite{fu2024differential}) and \emph{runahead execution} (e.g., VR~\cite{naithani2021vector} and DVR~\cite{naithani2023decoupled}) typically initiate chain discovery at a \emph{striding load} and define the speculated chain as the set of dependent loads reachable from that striding load. This design choice is convenient, but also restrictive: being dependent on a striding load is neither a necessary condition nor a sufficient condition for a load to be an irregular. In contrast, ICP directly treats irregular memory instructions as prefetching targets and does not constrain the starting point of dependency discovery to striding accesses.}

\mm{\textbf{Mechanisms for speculative execution.} \emph{Indirect prefetchers} typically rely on the ability of a stride prefetcher to predict the full future address of striding loads. This allows them to obtain the data accessed by the striding load from prefetched cache lines and use it to speculatively execute the chain. As a result, their speculative execution model is tightly coupled to stride-driven instruction chains. \emph{Runahead executions}, on the other hand, speculatively execute the dependency chains by leveraging spare core resources or another subthread. However, this capability comes at the cost of substantial hardware complexity, as it requires intrusive support in the CPU core to manage speculative execution. In contrast, ICP exploits cache-line responses associated with any existing prefetcher or demand request to trigger speculative execution, improving generality without incurring runahead-level complexity.}

ICP outperforms the state-of-the-art temporal prefetcher Triangel~\cite{ainsworth2024triangel} (14.0\% performance speedup) and indirect prefetcher DMP~\cite{fu2024differential} (6.0\% performance speedup) across both irregular SPEC CPU benchmarks~\cite{SPEC2006}, which represent recurring irregular memory access patterns and have been evaluated in previous temporal prefetching works~\cite{ainsworth2024triangel,wu2019efficient,wu2019temporal,wu2021practical}, and GAP benchmarks~\cite{beamer2015gap}, which primarily feature non-recurring indirect memory accesses and are widely used in indirect prefetching studies~\cite{fu2024differential,fu2025magellan,jamilan2022apt,talati2021prodigy}. \mm{ICP achieves these performance gains with only 449~B metadata storage and \mmi{1.7~KB} control state, significantly lower than temporal prefetchers, which require up to 1~MB of metadata storage and 17.6~KB control state.} Furthermore, our evaluations demonstrate that ICP achieves a substantially higher correlation reuse ratio than temporal prefetchers and exhibits greater correlation generality than indirect prefetchers.

\section{Background and Motivation}
\label{sec:background}


\subsection{Temporal Prefetching}
\label{subsec:tp}


Temporal prefetching has been proposed to address irregular memory access patterns that exhibit recurrence across time. The core idea of temporal prefetching \cite{jain2013linearizing, nesbit2004data, wu2019efficient, wenisch2009practical, wu2019temporal, bakhshalipour2018domino, ainsworth2024triangel, wu2021practical} is to record previously accessed memory addresses as metadata. When these addresses are accessed again, the prefetcher leverages the recorded metadata to predict subsequent memory accesses. 





\begin{figure}[t]
\centering
\includegraphics[width=.99\linewidth]{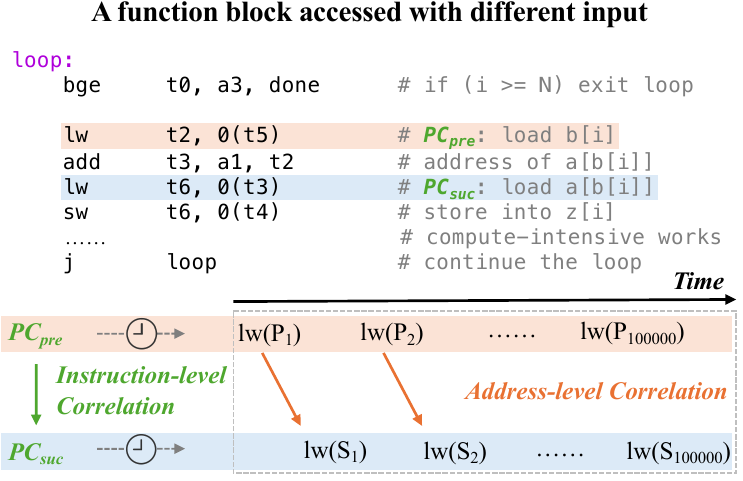}
\vspace{-.25in}
\caption{Comparison between instruction-level correlations and memory-address-level correlations. Instruction-level correlations efficiently capture the patterns, whereas address-level correlations fail to do so.}
\vspace{-.2in}
\label{fig:corr}
\end{figure}

Existing temporal prefetchers face fundamental challenges:

\textbf{Reliance on memory address repetition.} Temporal prefetchers operate at the memory address level, which makes them ineffective at capturing correlations between memory access instructions when their accessed addresses do not recur over time. Figure~\ref{fig:corr} illustrates this problem. In the example, two dependent memory access instructions within a loop are repeatedly executed. However, since the loop itself is executed with different inputs, all memory addresses generated by these instructions appear only once and never recur in the future. As a result, applying temporal prefetchers to this pattern would fail to prefetch any memory access. Nevertheless, \emph{dependence pattern between these two instructions does repeat} across loop iterations, yet temporal prefetchers fail to capture it.

\textbf{Substantial metadata storage.} All temporal prefetchers rely on large metadata structures to record correlations between memory addresses. Due to this significant capacity requirement, early designs placed metadata storage in off-chip DRAM. More recently, state-of-the-art prefetchers such as Triage~\cite{wu2019temporal,wu2021practical} and Triangel~\cite{ainsworth2024triangel} co-locate metadata storage with the on-chip Last-Level Cache (LLC), thereby eliminating the need to fetch metadata from off-chip memory. However, this approach comes at a high cost: according to Triage and Triangel, the metadata storage can occupy up to half of the LLC capacity (i.e., 1~MB). To mitigate this overhead, prior work~\cite{alecto,prophet} has proposed techniques for managing metadata storage more efficiently, including improved insertion and replacement policies as well as dynamic resizing. Nevertheless, even with these optimizations, metadata storage still requires hundreds of KB. This substantial requirement not only consumes valuable on-chip resources but also increases hardware complexity (e.g., enforcing way partitioning between the normal LLC and metadata storage).


\subsection{Indirect Prefetching}

Indirect prefetching~\cite{fu2024differential, talati2021prodigy, zhang2024rpg2, jamilan2022apt, yu2015imp, ainsworth2016graph, ainsworth2019software} has been proposed to address nested array accesses, such as $a[b[i]]$. Indirect prefetchers typically operate in two stages: identification and prefetching. \mm{In the identification stage, they detect the innermost array access (e.g., $b[i]$) and its dependent outer array accesses (e.g., $a[b[i]]$). Since such nested array patterns are usually enclosed within loops, the innermost array access often exhibits regular stride behavior, which can be effectively captured by a stride prefetcher. Indirect prefetchers then monitor the data produced by the striding load and correlate these values with the addresses accessed by the dependent load to identify the outer array access. In the prefetching stage, indirect prefetchers leverage the data prefetched by the stride prefetcher ($b[i+d]$) together with the base address of the outer array to generate prefetch requests.} 

\mm{The primary limitation of indirect prefetchers lies in their ineffectiveness in handling memory access patterns beyond nested array accesses, as both their identification and prefetching mechanisms are tightly coupled to array semantics. In particular, indirect prefetchers assume that $\textit{PC}_\text{inner}$ (the load accessing the innermost array) corresponds to a striding load—even in Tyche~\cite{xue2024tyche}, which aims to capture more general indirect patterns. This assumption provides several conveniences: (1) the indirection relationship can be identified by first detecting a striding load as $\textit{PC}_\text{inner}$ and then correlating the data it produces with addresses accessed by dependent loads; and (2) a stride prefetcher can predict the full memory address of $\textit{PC}_\text{inner}$, allowing the line offset to be used to accurately extract accessed value from the prefetched cache line. However, once $\textit{PC}_\text{inner}$ does not exhibit regular stride accesses (e.g., \textit{if (condition) x = a[b[i]].}), indirect prefetchers fail to identify and prefetch for such dependency patterns.}


\subsection{Runahead-based Schemes}

\mm{Runahead-based schemes accelerate irregular workloads by speculatively pre-executing future instructions to compute and issue memory accesses ahead of demand. Vector Runahead (VR)~\cite{naithani2021vector} extends classic runahead by speculatively vectorizing future loop iterations into SIMD-style operations (e.g., gathers) so it can issue a batch of dependent misses in parallel and expose high MLP. Decoupled Vector Runahead (DVR) improves over VR by removing the key trigger limitation of conventional runahead: being constrained by the ROB-stall window. DVR decouples runahead from the main OoO pipeline and runs it in a subthread context, enabling more proactive lookahead and higher sustained MLP.}

\mm{Runahead-based schemes incur significant hardware complexity. Both VR and DVR require substantial CPU-side modifications to support speculative execution and vectorization. DVR further introduces a decoupled subthread context to execute runahead slices, making them far more invasive than cache-local prefetchers. Moreover, VR and DVR initiate dependency-chain discovery from striding loads like indirect prefetchers, which inherently restricts their generality.}

\subsection{Motivations}

Motivated by the fundamental limitations of existing temporal and indirect prefetchers, we introduce a new prefetching scheme for irregular memory accesses. The scheme works at the instruction level and is designed to capture general correlation patterns among \emph{irregular} but \emph{recurring} memory access instructions. We define two memory access instructions as \textbf{correlated} if a data-dependency path exists from one instruction to the other. Specifically, at runtime, we identify correlated and recurring memory instruction pairs ($\textit{PC}_\text{pre}, \textit{PC}_\text{suc}$), where $\textit{PC}_\text{suc}$ corresponds to irregular memory instructions, \mm{and $\textit{PC}_\text{pre}$ can be any memory instructions that producing data for $\textit{PC}_\text{suc}$.} The identified correlations are then maintained in a lightweight table. When $\textit{PC}_\text{pre}$ is re-executed after its correlations have been established, ICP leverages the execution results of $\textit{PC}_\text{pre}$ to speculatively execute the data-dependency path toward $\textit{PC}_\text{suc}$ and prefetch its corresponding memory accesses.

\textbf{Advantages over temporal prefetching.} Compared to temporal prefetching, our scheme works at the instruction level rather than the memory address level, offering the following advantages: (1) our scheme can capture certain patterns where temporal prefetching fall shorts. As shown in Figure~\ref{fig:corr}, even if the memory addresses accessed by two correlated instructions do not recur in the future, the correlation between these instructions still repeat. Our scheme leverages this instruction-level recurrence to issue prefetches in such cases. (2) Our scheme requires significantly less storage than temporal prefetchers, as it only maintains instruction-level correlation information in tables without tracking the individual memory addresses they generate. 


\textbf{Advantages over indirect prefetchers.} Our scheme can handle a much broader range of memory access patterns than indirect prefetching. We support any types of ($\textit{PC}_\text{pre}, \textit{PC}_\text{suc}$) pairs without restricting $\textit{PC}_\text{pre}$ and $\textit{PC}_\text{suc}$ to array access instructions. For example, $\textit{PC}_\text{pre}$ and $\textit{PC}_\text{suc}$ may correspond to pointer-based accesses, such as $*p$ and $(*p).next$. Alternatively, $\textit{PC}_\text{pre}$ and $\textit{PC}_\text{suc}$ may correspond to different types of accesses, such as an array-based access $a[i]$ and a pointer-based access $*a[i]$. In contrast, indirect prefetchers can only identify and prefetch a limited set of array-based pairs.

\textbf{Advantages over runahead-based scheme.} \mm{Compared to runahead-based approaches such as VR and DVR, ICP achieves substantially lower hardware complexity. Runahead schemes require non-trivial modifications to the CPU core to support speculative execution, vectorization, checkpointing, and recovery mechanisms. In contrast, ICP operates entirely within the cache hierarchy and interacts with the core in a non-intrusive manner, without modifying the CPU main pipeline. As a result, ICP provides better benefits for irregular memory accesses (Section~\ref{subsec:perf_eva}) while maintaining a significantly simpler, more modular, and easier-to-integrate hardware design.}

\section{Overview}
\label{sec:overview}

\begin{figure}[t]
\centering
\includegraphics[width=.99\linewidth]{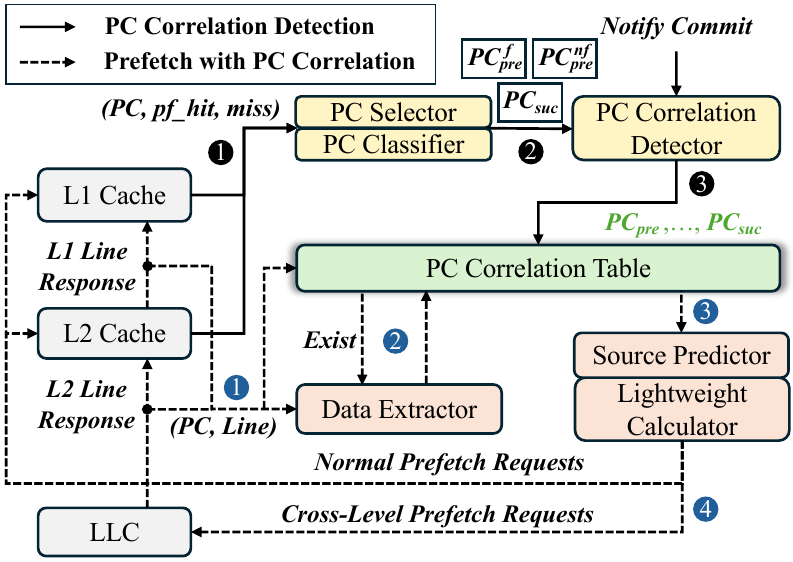}
\vspace{-.25in}
\caption{ICP Overview.}
\vspace{-.23in}
\label{fig:overview}
\end{figure}

In response to the motivations, we propose \textbf{ICP}, a new hardware prefetching scheme that records observed memory instruction correlations in a lightweight table and leverages this information to prefetch irregular memory accesses. Figure~\ref{fig:overview} illustrates the primary structures and workflow of ICP. The design includes two stages: \emph{PC correlation detection} and \emph{prefetching with PC correlations}.

\subsection{PC Correlation Detection}
\label{subsec:corr_det}

In this stage, we identify correlated memory-instruction pairs ($\textit{PC}_\text{pre}, \textit{PC}_\text{suc}$), where $\textit{PC}_\text{suc}$ issues irregular memory accesses. To facilitate this process, we introduce the following hardware components:




\textbf{PC Selector.} This component is used to select PCs that are likely candidates for either $\textit{PC}_\text{pre}$ or $\textit{PC}_\text{suc}$ (Step \ding{172}). 


\textbf{PC Classifier.} This component further classifies all $\textit{PC}_\text{pre}$ instructions into two categories: basic-prefetcher-friendly and non-basic-prefetcher-friendly. The rationale behind this classification is that the memory data of $\textit{PC}_\text{pre}$ can be obtained either from prefetch requests issued by basic hardware prefetchers or from demand accesses. Although both types of data can be used to execute the dependency path toward $\textit{PC}_\text{suc}$, they differ in their impact on prefetching timeliness and accuracy. Data obtained from prefetch requests can initiate the execution of the dependency path earlier than demand accesses, thereby improving prefetching timeliness for $\textit{PC}_\text{suc}$. However, inaccurate data generated by basic hardware prefetchers may ultimately lead to incorrect prefetches for $\textit{PC}_\text{suc}$. To balance timeliness and accuracy, ICP adopts two distinct prefetching schemes for these categories: (1) if $\textit{PC}_\text{pre}$ can be effectively handled by basic prefetchers, its prefetched data is primarily used to compute the memory address of $\textit{PC}_\text{suc}$, improving prefetching timeliness; and (2) if $\textit{PC}_\text{pre}$ cannot be effectively handled by basic prefetchers, ICP relies on the data from $\textit{PC}_\text{pre}$’s demand accesses to prefetch for $\textit{PC}_\text{suc}$, reducing inaccurate prefetches.

\textbf{PC Correlation Detector.} This component leverages the candidate sets of $\textit{PC}_\text{pre}$ and $\textit{PC}_\text{suc}$ provided by PC Selector and PC Classifier (Step \ding{173}) to identify correlated pairs ($\textit{PC}_\text{pre}, \textit{PC}_\text{suc}$). To establish correlations between $\textit{PC}_\text{pre}$ and its potential $\textit{PC}_\text{suc}$, the PC Correlation Detector monitors committed instructions and employs a simplified $O(N)$ register renaming mechanism to construct data-dependency trees rooted at $\textit{PC}_\text{pre}$ that extend to other $\textit{PC}_\text{suc}$ instructions. Each path from $\textit{PC}_\text{pre}$ to $\textit{PC}_\text{suc}$ defines a correlated instruction pair ($\textit{PC}_\text{pre}, \textit{PC}_\text{suc}$), consisting of a \textbf{$\boldsymbol{\textit{PC}_\text{pre}}$ node}, a \textbf{$\boldsymbol{\textit{PC}_\text{suc}}$ node}, and \textbf{intermediate instruction nodes} that form the dependency chain between them. 

\textbf{PC Correlation Table.} This component records all identified correlated instruction pairs (Step \ding{174}) and serves as the interface between the \emph{PC correlation detection} module and the \emph{prefetching with PC correlation} module. Compared to the metadata table used in temporal prefetching, the PC Correlation Table introduces significantly lower storage overhead (less than 1~KB), as it only needs to record instruction-level correlations rather than memory address correlations.

\subsection{Prefetching with PC Correlations}

ICP monitors cache line responses (Step \ding{172}, dotted in blue). When $\textit{PC}_\text{pre}$ is re-executed after its correlations have been identified, the correlations stored in the PC Correlation Table are used to prefetch the upcoming memory accesses of $\textit{PC}_\text{suc}$. To facilitate the prefetching process, we develop three hardware components: the Data Extractor, the Lightweight Calculator, and the Source Predictor.

\textbf{Data Extractor.} At Step \ding{173}, if the data response corresponds to a $\textit{PC}_\text{pre}$ entry recorded in the table, ICP employs Data Extractor to retrieve data accessed by $\textit{PC}_\text{pre}$ from the returned cache line. If $\textit{PC}_\text{pre}$ is identified as basic-prefetcher-friendly, ICP uses Data Extractor to predict the data within the cache line fetched by both prefetcher and demand accesses. Otherwise, ICP extracts data \emph{only} from cache lines fetched by demand accesses.

\textbf{Lightweight Calculator and Source Predictor.} At Step \ding{174}, the Lightweight Calculator uses the data retrieved by Data Extractor to speculatively execute the instruction dependent on $\textit{PC}_\text{pre}$. The dependent instruction may be an intermediate instruction along the dependency path or the $\textit{PC}_\text{suc}$ instruction itself. The Lightweight Calculator recursively executes the chained dependent instructions until the speculative execution reaches $\textit{PC}_\text{suc}$. When source registers fall outside the identified dependency path, the Source Predictor predicts their corresponding values. Finally, at Step \ding{175}, ICP computes the target memory address of $\textit{PC}_\text{suc}$ and generates the corresponding prefetch request. When the data of $\textit{PC}_\text{pre}$ is obtained from a demand-fetched line, the prefetch request is forwarded to the LLC to enhance prefetch timeliness. Otherwise, ICP issues the prefetch request to the current cache level.

\section{Hardware Design}
\label{sec:design}

\subsection{PC Selector and Classifier} 

\begin{figure}[t]
\centering
\includegraphics[width=.99\linewidth]{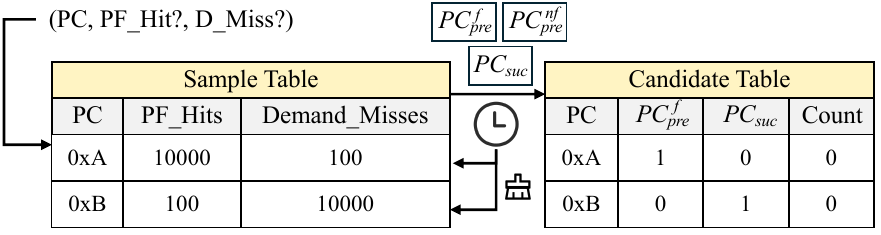}
\vspace{-.25in}
\caption{PC Selector and Classifier.}
\vspace{-.2in}
\label{fig:selector}
\end{figure}

The PC Selector and PC Classifier record performance counters of memory instructions to construct the $\textit{PC}_\text{pre}^\text{f}$ ($\textit{PC}_\text{pre}$ and basic-prefetcher-friendly) candidate set, $\textit{PC}_\text{pre}^\text{nf}$ ($\textit{PC}_\text{pre}$ and non-basic-prefetcher-friendly) candidate set, and the $\textit{PC}_\text{suc}$ candidate set. For each demand request arriving at ICP, three types of information are extracted: its PC address, whether it results in a prefetch hit, and whether it results in a demand miss. As shown in Figure~\ref{fig:selector}, ICP maintains a Sample Table to track the performance counters. The Sample Table is indexed by the PC address of memory access instructions. When a PC experiences a prefetch hit or a demand miss, ICP increments the corresponding counter value in the Sample Table. To retain frequently accessed PCs, the Sample Table is managed using an LRU replacement policy. Due to distinct memory access patterns at different cache levels, ICP maintains separate instances of the Sample Table for the L1 and L2 caches.

The Sample Table periodically exports the $\textit{PC}_\text{pre}$ and $\textit{PC}_\text{suc}$ candidates for subsequent uses. ICP defines an epoch $e$ as the number of times the Sample Table has been accessed. At the end of each epoch, ICP ranks all PCs based on their demand miss counts and evaluates the prefetching coverage of each PC. Specifically, ICP determines whether a PC qualifies as a $\textit{PC}_\text{suc}$ candidate using Equation~\ref{eq:suc}:


\begin{equation}
\label{eq:suc}
\mathcal{S}_{suc} = \text{Top}_{n}\big(\{\, PC_i \mid \text{misses}(PC_i) \ge \theta_{\text{miss}} \,\}\big)
\end{equation}

The formula selects $n$ PCs with the highest demand miss counts as $\textit{PC}_\text{suc}$ candidates, where $n$ is a design parameter controlled by the system designer. \mm{The rationale behind Equation~\ref{eq:suc} is that, in a system already equipped with basic prefetchers, PCs that still incur the most cache misses are likely not well served by those prefetchers and therefore are more likely to correspond to irregular memory accesses.} ICP adjusts both the epoch length and the Sample Table size to ensure that the collected PCs accumulate sufficient demand misses for accurate classification. Specifically, prolonging the epoch length allows the Sample Table to gather enough statistical information, while restricting the Sample Table size ensures that PCs with fewer demand misses are gradually evicted and replaced by more memory-intensive ones. Additionally, ICP uses condition $\text{misses}(PC_i) \ge \theta_{\text{miss}}$ to filter out PCs with insignificant misses. 


Then, ICP uses Equation~\ref{eq:pre} to determine whether a PC is selected as a $\textit{PC}_\text{pre}^\text{f}$:

{\small
\begin{equation}
\label{eq:pre}
\left\{
\begin{aligned}
\text{Cov}(PC_i) &=
\frac{\text{PF\_Hits}(PC_i)}
{\text{PF\_Hits}(PC_i)+\text{Demand\_Misses}(PC_i)}, \\
\mathcal{S}_{pre}^{f} &=
\text{Top}_{n}\!\left(
\left\{\, PC_i \mid \text{Cov}(PC_i) \ge \theta_{\text{cov}} \,\right\}
\right)
\end{aligned}
\right.
\end{equation}
}

The formula selects $n$ PCs with the highest prefetching coverage as $\textit{PC}_\text{pre}^\text{f}$ candidates. \mm{They could serve the potential data producer for $\textit{PC}_\text{suc}$.} Similar to Equation~\ref{eq:suc}, ICP applies a threshold parameter $\theta_{\text{cov}}$ to filter out PCs with low prefetching coverage. In our experiments, we set $\theta_{\text{cov}} = 0.1$.

ICP leverages an important observation to construct the candidate set $\mathcal{S}_{pre}^{nf}$: \textbf{a $\boldsymbol{\textit{PC}_\text{suc}}$ can also serve as a $\boldsymbol{\textit{PC}_\text{pre}}$}. Consider the nested pointer access pattern $*(*(*p))$. There are three memory instructions, $\textit{PC}_\text{1}$, $\textit{PC}_\text{2}$, and $\textit{PC}_\text{3}$, that access the pointers at $p$, $*p$, and $*(*p)$, respectively. $PC_1$ is regular because it originates from an array of pointer structures, and therefore serves as the $\textit{PC}_\text{pre}$ for $\textit{PC}_\text{2}$. $\textit{PC}_\text{2}$, which exhibits irregular behavior, acts as a $\textit{PC}_\text{suc}$ dependent on $\textit{PC}_\text{1}$. Furthermore, $\textit{PC}_\text{2}$ can also serve as $\textit{PC}_\text{pre}$ for the subsequent instruction $\textit{PC}_\text{3}$. Driven by this insight, ICP directly reuses the set of $\mathcal{S}_{suc}$ as $\mathcal{S}_{pre}^{nf}$, as defined in Equation~\ref{eq:npre}:

\begin{equation}
\label{eq:npre}
\mathcal{S}_{pre}^{nf} = \mathcal{S}_{suc}
\end{equation}

Then, as shown in Figure~\ref{fig:selector}, ICP employs another structure, Candidate Table (duplicated for L1 and L2 cache like Sample Table), to store identified $\mathcal{S}_{suc}$, $\mathcal{S}_{pre}^{f}$, and $\mathcal{S}_{pre}^{nf}$. Since the content of $\mathcal{S}_{suc}$ is identical to that of $\mathcal{S}_{pre}^{nf}$, ICP reduces the storage overhead by compressing Candidate Table and using a single field to indicate whether a PC entry belongs to $\textit{PC}_\text{pre}^\text{nf}$ or $\textit{PC}_\text{suc}$. The Candidate Table serves as the interface between PC Selector/Classiﬁer and PC Correlation Detector. After storing these three sets into Candidate Table, ICP clears Sample Table and continues searching for potential $\textit{PC}_\text{pre}$ and $\textit{PC}_\text{suc}$ instructions in the next execution phase.

\subsection{PC Correlation Detector}
\label{subsec:detector}

The PC Correlation Detector leverages the candidate sets stored in Candidate Table to identify correlated instruction pairs ($\textit{PC}_\text{pre}$, $\textit{PC}_\text{suc}$). To discover these correlations, ICP  constructs data-dependency trees rooted at each $\textit{PC}_\text{pre}$, which expands toward its dependent $\textit{PC}_\text{suc}$ instructions. Each path from a $\textit{PC}_\text{pre}$ node to a $\textit{PC}_\text{suc}$ node represents a correlated instruction pair ($\textit{PC}_\text{pre}$, $\textit{PC}_\text{suc}$).


\subsubsection{Dependency Tree Construction Triggering} 


The PC Correlation Detector leverages the committed instruction information from the ROB to construct dependency trees. The construction process is triggered whenever the PC of a committed instruction belongs to either $\textit{PC}_\text{pre}^\text{f}$ or $\textit{PC}_\text{pre}^\text{nf}$. Once a dependency tree construction is initiated, ICP temporarily blocks subsequent construction requests, even if they correspond to another $\textit{PC}_\text{pre}$. This blocking mechanism prevents resource contention. To avoid a scenario where a single $\textit{PC}_\text{pre}$ monopolizes hardware resources and causes deadlock, ICP introduces a \texttt{Count} field in the Candidate Table. This field records the number of attempts made to construct the dependency tree for the corresponding PC. When the count exceeds a predefined threshold, the PC Correlation Detector ceases further construction attempts for that PC.

\begin{figure}[t]
\centering
\includegraphics[width=.99\linewidth]{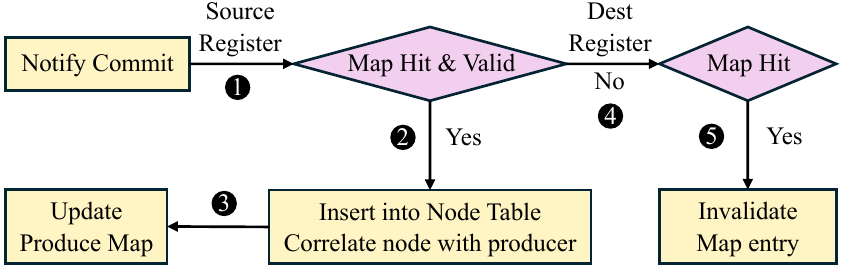}
\vspace{-.2in}
\caption{Process of dependency tree construction.}
\vspace{-.2in}
\label{fig:corr_process}
\end{figure}

\begin{figure}[t]
\centering
\includegraphics[width=.99\linewidth]{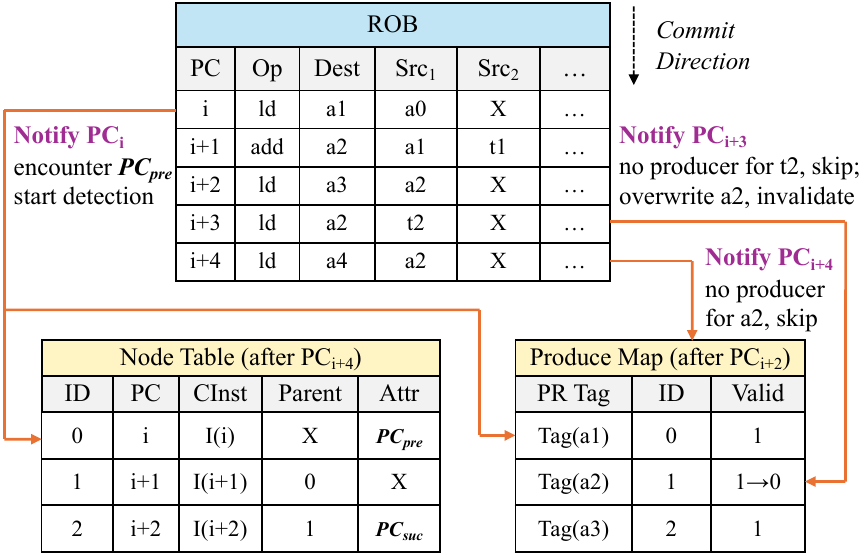}
\vspace{-.25in}
\caption{Example of dependency tree construction.}
\vspace{-.25in}
\label{fig:corr_iden}
\end{figure}

\subsubsection{Dependency Tree Construction} 

As shown in Figure~\ref{fig:corr_iden}, ICP employs two hardware structures: the Node Table and the Produce Map, to construct the dependency tree rooted at the triggered $\textit{PC}_\text{pre}$. The \textbf{Node Table} primarily records correlations between different PCs within the dependency tree (through the \texttt{ID} and \texttt{Parent} fields) and stores auxiliary information that enables reconstruction of the instruction-level dependency path from $\textit{PC}_\text{pre}$ to $\textit{PC}_\text{suc}$ (through the \texttt{PC}, \texttt{CInst}, and \texttt{Attr} fields). By hitchhiking on the ROB commit process, ICP adopts the core concept of register renaming and implements a simplified $O(N)$ register renaming scheme to efficiently track data dependencies among committed instructions during dependency tree construction. The \textbf{Produce Map} maintains the mapping between physical registers and their producer PC nodes. Specifically, it tracks which PC node last produced the value for each physical register (\texttt{PR Tag}), represented by the \texttt{ID} field in the Node Table, and uses the \texttt{Valid} field to indicate whether that mapping is currently valid. We now describe the process of dependency tree construction as follows:

First, upon triggering dependency tree construction, ICP initializes the Node Table and Produce Map using the committed instruction at the head of the ROB. ICP allocates a new entry in the Node Table with ID = 0 and records its compressed instruction format (detailed in Section~\ref{subsec:recording}), PC address, and attribute, which indicates whether the PC belongs to $\textit{PC}_\text{pre}^\text{f}$, $\textit{PC}_\text{pre}^\text{nf}$, or $\textit{PC}_\text{suc}$. Then, as shown in Figure~\ref{fig:corr_process} and Figure~\ref{fig:corr_iden}, ICP updates the Node Table and Produce Map using information from subsequently committed instructions.

\begin{enumerate}
    \item \textbf{Checking Source Registers.} Each source register of the committed instruction notified from the ROB is sent to the Produce Map for lookup.
    \item \textbf{Updating Node Table.} If any source register hits in the Produce Map, the committed instruction is determined to be data-dependent on the producer instruction indicated by the corresponding \texttt{ID} entry in the Produce Map (as illustrated by $PC_{i+1}$ and $PC_{i+2}$ in Figure~\ref{fig:corr_iden}). In such case, ICP inserts a new entry at the end of the Node Table and increments the node \texttt{ID} by one. It sets the \texttt{Parent} field of the newly inserted node to the node \texttt{ID} of its producer instruction. If the instruction belongs to $\textit{PC}_\text{suc}$, ICP updates its \texttt{Attr} field accordingly.
    \item \textbf{Updating Produce Map.} ICP updates the Produce Map using the destination register of the committed instruction. The \texttt{ID} field in the Produce Map is set to the corresponding \texttt{ID} previously allocated in the Node Table, and the \texttt{Valid} field is initialized to 1.
    \item \textbf{Checking Destination Register.} If no correlation with previous instructions is found for the current committed instruction, ICP does not maintain any information about this instruction in either the Node Table or the Produce Map ($PC_{i+3}$ and $PC_{i+4}$ in Figure~\ref{fig:corr_iden}). 
    \item \textbf{Invalidating Map Entry.} Although ICP skips committed instructions without correlations, such instructions may overwrite physical registers whose values were produced by earlier instructions already tracked in the Produce Map (e.g., $PC_{i+3}$). In these cases, ICP invalidates the corresponding entry in the Produce Map to maintain the correctness of data-dependence correlations.
\end{enumerate}

ICP defines two termination conditions for the above process. First, ICP limits the maximum number of committed instructions processed during each dependency tree construction trigger. Our experiments set this limit to 128. Second, ICP restricts the size of Node Table. When the number of recorded nodes reaches the maximum capacity of Node Table, ICP terminates the dependency tree construction for current $\textit{PC}_\text{pre}$. Our experiments set Node Table size to 16. The above two termination conditions are combined using a logical OR.

\subsubsection{Correlated Instruction Pairs Reconstruction} 
\label{subsubsec:generation}

ICP traverses the \emph{Node Table} to identify and generate correlated instruction pairs ($\textit{PC}_\text{pre}$, $\textit{PC}_\text{suc}$). With a Node Table capacity of $N = 16$, reconstructing all dependency paths from the root ($\textit{PC}_\text{pre}$, entry $\texttt{ID} = 0$) to every $\textit{PC}_\text{suc}$ requires:

\begin{enumerate}
    \item a single linear scan of the table to collect the $\textit{PC}_\text{suc}$ entries ($\le N - 1$); and
    \item for each $\textit{PC}_\text{suc}$ entry, following its \texttt{Parent} pointers back to $\texttt{ID} = 0$.
\end{enumerate}

The total computational effort corresponds to the aggregate of all path lengths from each $\textit{PC}_\text{suc}$ node to the root. In the worst-case scenario, this overhead can be expressed as:

\[
\sum_{d=1}^{N-1} d = \frac{N(N - 1)}{2}
\]

When $N = 16$, we require at most 120 parent-pointer dereferences. Assuming a 1-cycle SRAM access per table lookup, the worst-case traversal requires no more than 120 cycles and utilizes only a small temporary buffer (up to 16 \texttt{ID}s) to record the path.

In practice, the dependency trees are typically shallow (Figure~\ref{fig:dep_dis}). By performing a one-time depth-first search (DFS) from the root node ($\textit{PC}_\text{pre}$) to all reachable $\textit{PC}_\text{suc}$ nodes, ICP reduces the total number of table reads to $O(N)$ (i.e., at most 16 cycles for 16 table lookups) to reconstruct all paths. Consequently, the path reconstruction overhead is negligible relative to overall program execution time.


\subsection{PC Correlation Recording}
\label{subsec:recording}

\begin{figure}[t]
\centering
\includegraphics[width=.99\linewidth]{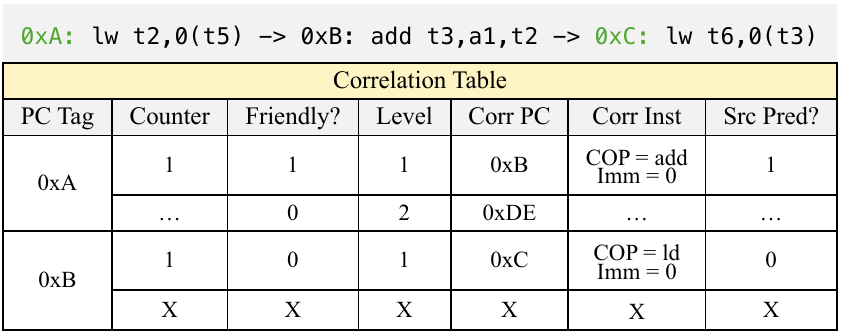}
\vspace{-.25in}
\caption{Correlation Table structure and usage.}
\vspace{-.23in}
\label{fig:corr_table}
\end{figure}

ICP employs the Correlation Table to record the instruction correlations identified in Section~\ref{subsec:detector}. Figure~\ref{fig:corr_table} illustrates the table structure and its usage. For each PC entry, ICP stores up to two successor instructions that consume the value produced by this PC. This choice reflects two realities: (i) a producer instruction may have multiple consumers, and (ii) program phase behavior can cause different successors to be observed across execution epochs. Consequently, retaining more than one successor per PC increases prefetching coverage while keeping storage overhead bounded. To handle cases where a PC has more than two successor instructions, ICP employs the \texttt{Counter} field in the Correlation Table to retain the most frequently observed successors. When a new successor is identified and all successor slots are occupied, the entry with the smallest \texttt{Counter} value is replaced. The \texttt{Friendly} field indicates whether the recorded PC is classified as basic-prefetcher-friendly. This information is later utilized during the prefetching phase for selecting cache line responses (from prefetches or demand accesses). The \texttt{Level} field associates each correlated instruction pair with its corresponding cache level. Since ICP maintains separate instances of the Sample Table and Candidate Table for different cache levels, this field is set to the cache level from which the correlated instruction pair is derived. During the prefetching phase, ICP uses the cache line response from the cache level indicated by the \texttt{Level} field to speculatively execute the successor. The \texttt{Corr\ PC} field stores the PC address of the successor instruction. The \texttt{Corr\ Inst} field records the essential information of the successor instruction, including the operation type supported by the Lightweight Calculator in ICP and its corresponding immediate value. ICP excludes any dependency paths containing operation types that are not supported by the Lightweight Calculator. Furthermore, a correlated instruction may contain multiple source registers, some of which may lie outside the identified dependency path. To enable speculative execution of such instructions during the prefetching phase, ICP employs the Source Predictor to estimate the values of these external source registers. The \texttt{Src\ Pred} field indicates whether the execution of a given correlated instruction requires this prediction mechanism.

\subsection{Prefetching with PC Correlations}
\label{subsec:pf_with_pcc}

During the prefetching phase, ICP monitors cache-line responses originating from both prefetch and demand requests. If a response corresponds to a PC entry recorded in the Correlation Table (excluding prefetched lines associated with non–basic-prefetcher-friendly PCs), ICP utilizes the data contained in that cache line to speculatively execute the successor instruction associated with the PC and ultimately issue prefetch requests for the $\textit{PC}_\text{suc}$. To support this process, ICP integrates three specialized components: the \textbf{Data Extractor}, which retrieves the data accessed by the PC within the cache line; the \textbf{Lightweight Calculator}, which performs speculative execution of the correlated successor instructions; and the \textbf{Source Predictor}, which predicts the values of source registers when a successor instruction contains operands outside the identified data-dependency path.

\textbf{Data Extractor} applies different data extraction methods depending on whether the cache line is fetched by a demand request or a prefetch request. If the cache line is fetched by a demand request, Data Extractor directly utilizes offset bits of the memory access address to locate and extract the target data from the cache line. Conversely, if the cache line is fetched by a prefetch request, Data Extractor must infer the offset of the target data within the line, as most cache prefetchers operate at the cache-line granularity and prefetch requests record only the line address without fine-grained offset information. To handle such cases, Data Extractor maintains a table that records historical line offsets observed from demand requests, which are then used to infer the offset for prefetched cache lines. Specifically, when a basic-prefetcher-friendly PC is added to Correlation Table, ICP simultaneously allocates an entry in Data Extractor. The Data Extractor associates each observed line offset with a counter to track its occurrence frequency. During the prefetching phase, it computes the relative probability of each offset by dividing its individual counter value by the sum of all counters across offsets. When the probability of a line offset exceeds a predefined threshold (set to 0.1 in our experiments), Data Extractor uses the corresponding offset to extract the target data from the prefetched cache line. 

\mm{The Data Extractor enables ICP to trigger speculation from cache-line responses that arrive \emph{far ahead} of $\textit{PC}_\text{pre}$, improving timeliness relative to indirect prefetchers. This is because indirect prefetchers typically rely on a stride prefetcher to predict the \emph{exact} address and extract the accessed word. With the same prefetch degree, a line-level prefetcher can generally fetch cache lines farther ahead than an exact-address stride prefetcher. Data Extractor allows ICP to utilize cache-line responses generated by \emph{any} existing prefetcher as valid triggers, enabling earlier and more general prefetches.}

\begin{table}[t]
\centering
\small
\setlength{\tabcolsep}{6pt}
\renewcommand{\arraystretch}{1.2}
\caption{\mm{Operations supported by the Lightweight Calculator.}}
\vspace{-.05in}
\label{tab:calculator_ops}
\begin{tabular}{|c|c|c|}
\hline
\textbf{Category} & \textbf{Operation} & \textbf{Example} \\
\hline
\multirow{4}{*}{Arithmetic}
 & ADD  & $x = a + b$ \\
 & SUB  & $x = a - b$ \\
 & SHL  & $x = a \ll k$ \\
 & SHR  & $x = a \gg k$ \\
\hline
\multirow{3}{*}{Logical}
 & AND  & $x = a\ \&\ b$ \\
 & OR   & $x = a\ |\ b$ \\
 & XOR  & $x = a \oplus b$ \\
\hline
\end{tabular}
\vspace{-.25in}
\end{table}

\textbf{Lightweight Calculator} leverages the data provided by the Data Extractor and the successor instruction information stored in the Correlation Table to speculatively execute the dependency path starting from $\textit{PC}_\text{pre}$. When the successor instruction is a memory access instruction, the Lightweight Calculator computes its target memory address and issues the corresponding prefetch request. Otherwise, the Lightweight Calculator recursively continues speculative execution along the dependency path. It is designed to support basic arithmetic and logical operations (Table~\ref{tab:calculator_ops}), as intermediate instructions between two correlated memory instructions are typically simple address-calculation operations. Our extensive evaluation results (Section~\ref{sec:eval}) confirm that this operation set is sufficient to achieve high performance. Consequently, the Lightweight Calculator introduces limited hardware complexity.

\textbf{Source Predictor} is employed to predict source register values when speculatively executing successor instructions with the \texttt{Src\ Pred} flag set. Since the intermediate instructions between $\textit{PC}_\text{pre}$ and $\textit{PC}_\text{suc}$ are memory address computation operations, the external source registers involved typically hold stable base addresses. Leveraging this property, Source Predictor leverages historical source register values to perform value prediction. Specifically, when a PC marked as \texttt{Src\ Pred} is added to the Correlation Table, ICP simultaneously allocates a corresponding entry in Source Predictor. Each entry records the PC address and the target source register to be predicted. During execution, Source Predictor monitors ROB to track committed values of corresponding source registers. It maintains a single recorded value for each tracked source register, along with a confidence bit that indicates prediction stability. When a newly observed value matches the recorded one, the confidence bit is set; otherwise, it is reset. The Source Predictor produces a prediction only when the confidence bit is set, and the predicted value is then used for speculative execution of the successor instructions.


\subsection{Integration with CPU and Memory Hierarchy}
\label{subsec:integration}

\mm{According to Figure~\ref{fig:overview}, ICP interfaces with both the CPU core and the memory hierarchy and relies on three categories of information: (1) demand requests, (2) cache-line responses returned from the memory hierarchy, and (3) committed instruction information from the CPU core. Since receiving demand requests is already a standard capability of hardware prefetchers, we focus primarily on how ICP accepts cache-line responses and committed instructions. }

\mm{Acquiring cache-line fill information, including both the returned data and the associated PC of the triggering memory instruction, is a common requirement in existing indirect prefetchers~\cite{fu2024differential,xue2024tyche,yu2015imp} and other cache-related microarchitecture designs~\cite{wu2011ship,subramaniam2009criticality}. Basically, returned data can be obtained by snooping the data bus~\cite{fu2024differential}. To propagate the associated PC, we extend each MSHR target entry with a compressed PC field, which is returned alongside the filled data. For compression, we observe that instructions within the same dependency chain often share identical high-order bits in their PC addresses. Therefore, we retain only the lower 3–4 bits of each PC and hash the high-order bits. Consequently, each compressed PC address requires only 10 bits. The storage overhead of this PC plumbing is: $N_{\text{MSHR}} \times T \times \text{PC}_{\text{bits}}$, where $N_{\text{MSHR}}$ is the number of MSHRs and $T$ is the number of targets per MSHR. For example, with $N_{\text{MSHR}} = 16$, $T = 8$, and a 10-bit PC, the additional storage is $16 \times 8 \times 10 = 1280$ bits, i.e., 160~B. This overhead is minimal.}

\mm{Similarly, leveraging committed instruction information is not unprecedented in prefetcher designs~\cite{ainsworth2021ghostminion,gerogiannis2023micro, ros2021entangling}. Importantly, ICP imposes very loose timing requirements on commit information because (1) the entire dependency chain construction process operates outside the CPU’s critical execution path (Section~\ref{subsec:detector}); (2) Once the dependency path between $(PC_{\text{pre}}, PC_{\text{suc}})$ is learned, it is recorded and reused for subsequent executions. Any delay in completing path construction merely postpones the issuance of prefetches for $PC_{\text{suc}}$ by a limited number of cycles. This loose timing constraint allows a clean decoupled interface. A small FIFO buffer can be inserted between the commit logic and ICP to asynchronously stream committed instruction information. The commit logic enqueues commit records, while ICP dequeues them at its own pace to construct dependency trees.}
\section{Evaluation}
\label{sec:eval}

\subsection{Experimental Setup}
\label{subsec:setup}

\begin{table}[h]
\vspace{-5mm}
\centering
\small
\caption{System Configuration.}
\vspace{-3mm}
\label{tab:config}
\resizebox{0.99\columnwidth}{!}{%
\begin{tabular}{|l|l|}
\hline
\textbf{Module} & \textbf{Configuration} \\
\hline
\hline 
Core &5-\mm{wide} fetch, 5-\mm{wide} decode \\
&10-\mm{wide} issue, 10-\mm{wide} commit \\
&\mm{120-entry IQ, 85/90-entry LQ/SQ} \\
& \mm{288-entry ROB, L-TAGE branch predictor}\\
\hline
Private L1 I/D cache & 64~KB each, 4-way, 64B line, 16 MSHRs  \\
&PLRU, 2 cycles hit latency \\
&Stream, stride, and spatial prefetchers \\ &scheduled by Alecto~\cite{alecto} for L1D cache \\
\hline
Private L2 cache & 512~KB, 8-way, 64B line, 32 MSHRs \\
&PLRU, 9 cycles hit latency, inclusive \\
\hline
Shared L3 cache & 2~MB/core, 16-way, 64B line, 36 MSHRs \\
&CHAR \cite{chaudhuri2012introducing}, mostly\_exclusive \\
&20 cycles hit latency \\
\hline
Memory 
& LPDDR5\_5500\_1x16\_BG\_BL32 \\
\hline
\end{tabular}
}
\vspace{-3mm}
\end{table}

\textbf{System Configuration}. We evaluate ICP using the full-system (FS) mode of gem5~\cite{binkert2011gem5}, with baseline system parameters summarized in Table~\ref{tab:config}. \mm{The configuration is designed to closely follow the original Triangel~\cite{ainsworth2024triangel} implementation, with the remaining parameters chosen to approximate those of the Arm Cortex X2.} To emulate a realistic prefetching environment similar to that of commercial CPUs, we integrate a hybrid configuration comprising multiple basic hardware prefetchers. We deploy stream, stride, and spatial prefetchers from IPCP~\cite{pakalapati2020bouquet} at the L1 cache, and coordinate them using Alecto~\cite{alecto}, a state-of-the-art prefetcher selection algorithm. These baseline prefetchers effectively capture regular memory access patterns but remain ineffective for irregular accesses, which ICP and other evaluated prefetchers target.

\textbf{Evaluated Prefetchers}. We compare ICP against representative prefetchers targeting irregular memory access patterns, including the state-of-the-art \emph{temporal prefetcher} Triangel~\cite{ainsworth2024triangel}, the \emph{indirect prefetcher} Tyche~\cite{xue2024tyche} and DMP~\cite{fu2024differential}, \mm{the \emph{runahead scheme} Vector Runahead~\cite{naithani2021vector} and Decoupled Vector Runahead~\cite{naithani2023decoupled}}, and the integrated design DMP+Triangel. For fair comparison, ICP and indirect prefetchers are both integrated at the L1 cache level. Triangel is integrated at the L2 cache level, following its original design~\cite{ainsworth2024triangel}. \mm{Runahead-based schemes are integrated with the CPU main pipeline.} 




\begin{figure*}[t]
\centering
\includegraphics[width=0.99\linewidth]{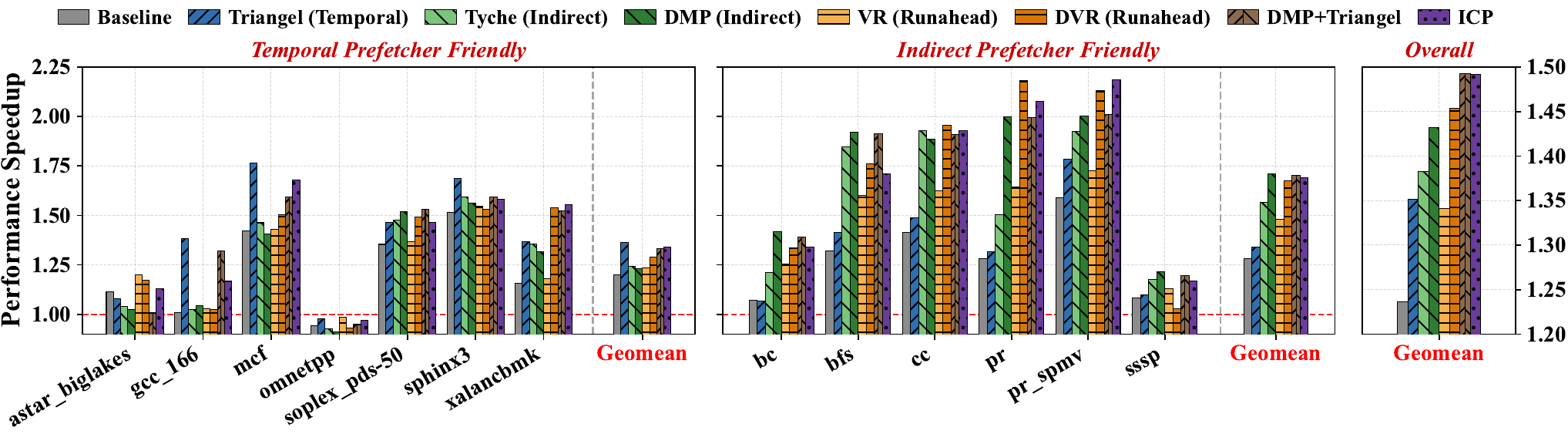}
\vspace{-.15in}
\caption{
\mm{IPC speedup on SPEC and GAP benchmarks. ICP outperforms Triangel by 13.99\%, Tyche by 10.86\%, DMP by 5.97\%, VR by 15.03\%, and DVR by 3.74\%. While ICP attains performance comparable to the combined DMP+Triangel configuration, it provides several key advantages: 1) Substantially smaller storage overhead; 2) Lower hardware complexity; 3) Lower DRAM traffic; 4) Higher energy efficiency.}}
\vspace{-.25in}
\label{fig:perf}
\end{figure*}

\textbf{Workloads}. We evaluate ICP using irregular SPEC CPU benchmarks~\cite{SPEC2006}, which are commonly used to assess existing temporal prefetchers~\cite{wu2019efficient, wu2019temporal, wu2021practical, ainsworth2024triangel}, and the GAP benchmark~\cite{beamer2015gap}, which is widely adopted by indirect prefetchers~\cite{jamilan2022apt, ainsworth2017software, fu2024differential, talati2021prodigy}. For GAP, we use the same input graphs as DMP~\cite{fu2024differential}. We apply the SimPoint technique \cite{sherwood2002automatically} to generate checkpoints across all workloads. Each \yao{SimPoint-sampled} checkpoint is warmed up with 200M instructions, followed by a simulation of the next 20M instructions. The reported metrics for each workload are calculated by aggregating the results from all its checkpoints with weighted averages.


\subsection{Performance Evaluation}
\label{subsec:perf_eva}

\begin{figure}[t]
\centering
\includegraphics[width=.99\linewidth]{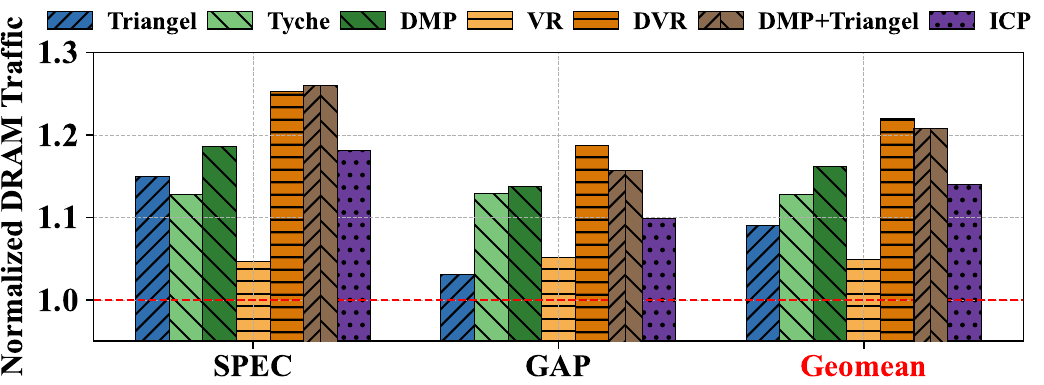}
\vspace{-.25in}
\caption{\mm{Comparison of DRAM traffic. ICP introduces 6.84\% less DRAM traffic than the DMP+Triangel hybrid configuration, 8.02\% than the DVR.}}
\vspace{-.2in}
\label{fig:dram}
\end{figure}

Figure~\ref{fig:perf} presents the performance speedup of ICP compared to other prefetchers designed for irregular memory access patterns. The evaluation spans two benchmark suites: SPEC, which features diverse and complex memory access patterns, and GAP, which primarily \mm{exhibits non-recurrent memory addresses but stable indirect instruction-level dependencies. The results demonstrate that ICP demonstrates consistent effectiveness across both SPEC and GAP benchmarks. ICP achieves a 25.51\% performance improvement over the baseline system equipped only with basic prefetchers, outperforming Triangel by 13.99\%, Tyche by 10.86\%, DMP by 5.97\%, VR by 15.03\%, DVR by 3.74\%, and performing on par with the combined DMP+Triangel configuration (25.61\%).} 

\mm{\textbf{Temporal prefetchers} perform well on SPEC benchmarks but achieve poor performance across most workloads in the GAP benchmark suite. In particular, Triangel struggles to capture irregular memory accesses that exhibit instruction-level repetition but lack explicit address-level repetition, which commonly occurs in GAP workloads (Section~\ref{subsec:icp_vs_triangel}).}

\mm{\textbf{Indirect prefetchers} are ineffective for many workloads in the SPEC benchmarks. Specifically, they fail to identify and prefetch more complex irregular access patterns that extend beyond nested memory accesses in SPEC (Section~\ref{subsec:icp_vs_dmp}).}

\mm{\textbf{Runahead-based schemes} deliver effective performance gains only when sufficient runahead time is available. Thus, VR fails to achieve strong performance across both SPEC and GAP. Although DVR improves over VR, it still struggles with workloads that exhibit complex dependency patterns, such as \textit{mcf} and \textit{gcc}. This limitation arises because DVR identifies dependency chains starting from striding loads, similar to indirect prefetchers. Moreover, a key drawback of runahead-based approaches is their substantial hardware complexity, as they require tight integration with the core pipeline (Section~\ref{subsec:icp_vs_runahead}).}

\mm{\textbf{Shortcomings of integrating indirect prefetchers with temporal prefetchers.} Although integrating DMP with Triangel achieves a performance gain comparable to ICP, this hybrid configuration inherits the limitations of both prefetching designs. \textbf{First}, the hybrid design incurs significantly greater hardware overhead than ICP. Specifically, Triangel and other temporal prefetchers typically require substantially larger metadata storage (up to 1MB), along with additional metadata management strategies (e.g., 17.6KB in Triangel~\cite{ainsworth2024triangel}). DMP additionally requires 912B storage. In contrast, ICP only requires 1.6~KB storage in total (Section~\ref{subsec:storage}). \textbf{Second}, the hybrid design introduces significant hardware complexity. For instance, to support temporal prefetchers, the system must share the LLC space with the metadata table, necessitating partitioning schemes and specific metadata table access, insertion, and replacement strategies. \textbf{Third}, the hybrid design causes higher DRAM traffic than ICP. Figure~\ref{fig:dram} illustrates the DRAM traffic for each evaluated prefetcher configuration, normalized to the baseline. The results indicate that ICP introduces a 13.98\% increase in DRAM traffic over the baseline system, compared to 9.04\% for Triangel, 16.21\% for DMP, and 20.82\% for the DMP+Triangel combination. Consequently, DMP+Triangel incurs 6.84\% more DRAM traffic than ICP. \textbf{Fourth}, the hybrid design is less energy efficient, consuming 4.9\% more enegy than ICP (Section~\ref{subsec:energy}).}

\subsection{ICP vs Temporal Prefetchers: More Efficient Correlations}
\label{subsec:icp_vs_triangel}

\begin{figure}[t]
\centering
\includegraphics[width=.99\linewidth]{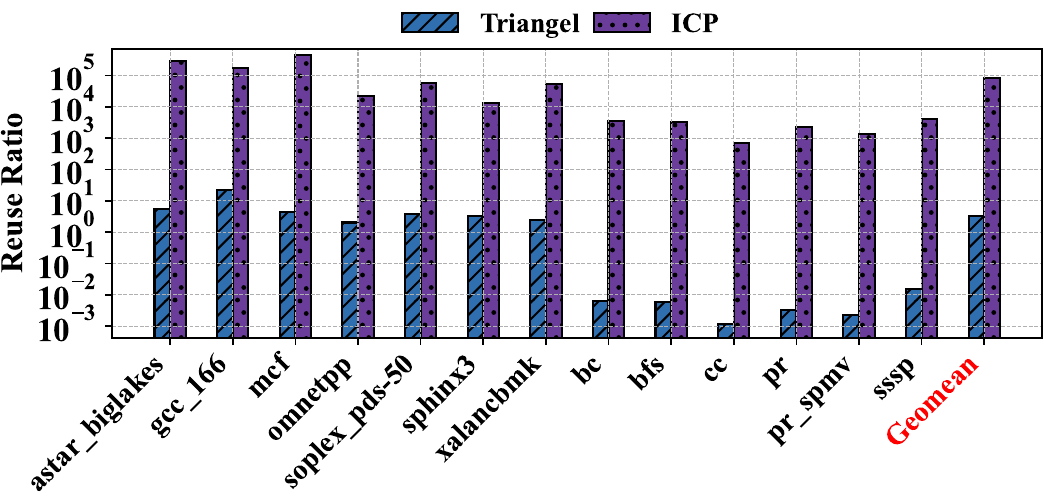}
\vspace{-.25in}
\caption{Comparison of metadata reuse ratio between ICP and Triangel. The reuse ratio in ICP is on the order of $10^5$ times higher than that of Triangel.}
\vspace{-.25in}
\label{fig:icp_vs_triangel}
\end{figure}

To demonstrate the efficiency of instruction-level correlations, Figure~\ref{fig:icp_vs_triangel} compares the metadata reuse ratio between ICP and Triangel. In this comparison, the metadata in ICP corresponds to instruction-level correlations, whereas in Triangel it represents address-level correlations. The reuse ratio is computed as the total number of metadata accesses divided by the number of metadata insertions. A higher reuse ratio indicates that the stored metadata is accessed more frequently after being recorded, reflecting higher metadata utilization.

The results show that the reuse ratio of metadata in ICP is on the order of $10^{5}$ times higher than that of Triangel. Combined with Figure~\ref{fig:figure1}, our evaluations demonstrate that instruction-level correlations are substantially more efficient than address-level correlations in capturing irregular memory access patterns. Furthermore, the metadata reuse ratio of Triangel in GAP benchmarks is extremely low, indicating that memory addresses in these workloads are rarely re-accessed. This observation aligns with the performance results shown in Figure~\ref{fig:perf}, where Triangel delivers limited benefits on GAP. In contrast, instruction-level correlations in them frequently reoccur, which validates ICP’s strong performance in GAP.

\subsection{ICP vs Indirect Prefetchers: More General Correlations}
\label{subsec:icp_vs_dmp}

\begin{figure}[t]
\centering
\includegraphics[width=.99\linewidth]{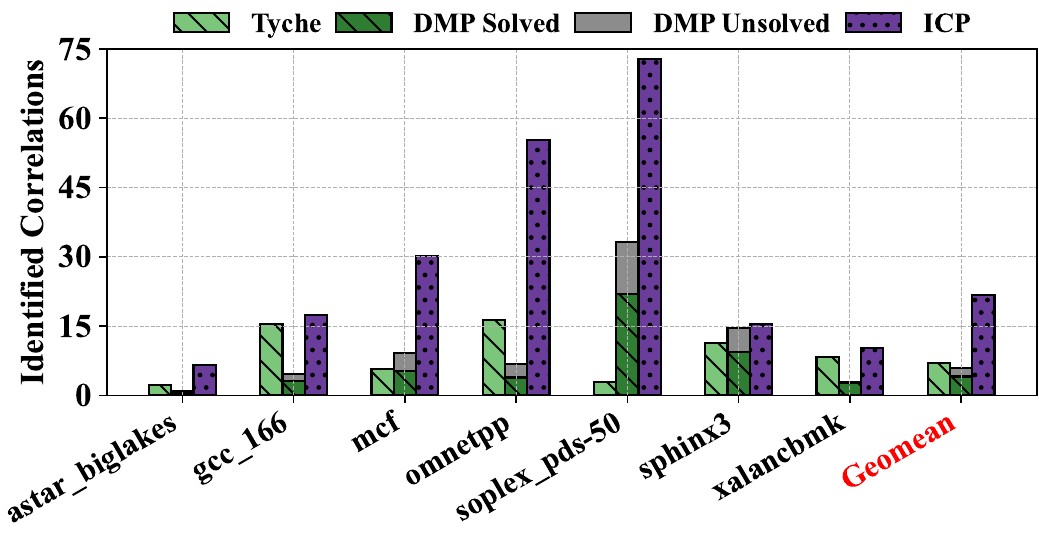}
\vspace{-.25in}
\caption{\mm{Comparison of identified instruction correlations between ICP and indirect prefetcher. ICP can identify and handle more general correlations.}}
\vspace{-.23in}
\label{fig:icp_vs_dmp}
\end{figure}

To demonstrate the broader generality of ICP compared to indirect prefetchers, Figure~\ref{fig:icp_vs_dmp} compares the total number of identified correlations across SPEC benchmarks, where \mm{Tyche and DMP} exhibits limited effectiveness (Figure~\ref{fig:perf}). The correlations identified by ICP represent general instruction-level relationships $(\textit{PC}_\text{pre}, \textit{PC}_\text{suc})$. \mm{In contrast, the correlations identified by Tyche correspond to \emph{IMA} patterns, while those in DMP correspond to \emph{indirection pairs}~\cite{fu2024differential}, both of which are specifically designed to capture data-dependence relationships between two nested array access instructions.}

We find that DMP’s differential matching mechanism~\cite{fu2024differential} can occasionally capture data dependencies between memory instructions beyond strictly nested array accesses, such as data dependencies in array-of-pointers structures (e.g., $p[i] \rightarrow *p[i]$). However, not all such patterns can be effectively handled by DMP, since its prefetching mechanism is explicitly designed for nested array structures. To highlight this limitation, Figure~\ref{fig:icp_vs_dmp} categorizes the indirection pairs identified by DMP into two groups: \emph{solved} and \emph{unsolved}, representing whether DMP can successfully prefetch for them.

\mm{Our results show that ICP identifies substantially more correlations than indirect prefetchers across the SPEC benchmarks, with an average of 22 correlations, compared to 7 for Tyche and 6 for DMP. Furthermore, nearly one-third (2 out 6) of the indirection pairs identified by DMP are ineffective. These findings underscore ICP's effectiveness in handling complex and diverse irregular memory access patterns. }


\subsection{ICP vs Runahead: Lower Hardware Complexity}
\label{subsec:icp_vs_runahead}

\begin{table*}[t]
\centering
\small
\caption{\mm{Integration complexity comparison between ICP and Runahead schemes.}}
\vspace{-.1in}
\label{tab:integration_icp_vs_dvr_box}
\begin{tabular}{|c|c|c|}
\hline
\textbf{Integration aspect} & \textbf{ICP (this work)} & \textbf{DVR} \\
\hline
\hline

\textbf{Thread requirement} &
Not required. &
Required. \\
\hline

\textbf{Decode} &
No changes. &
Mode-aware (Discovery or DVR) decode path.
\\
\hline

\textbf{Execute} &
No changes. &
Support for DVR-generated vectorized address-generation ops. \\
\hline

\textbf{Commit} &
Asynchronous Buffer. &
\makecell[c]{Ensure DVR-generated ops not update architectural state \\ Termination/recovery logic When DVR stops}
\\
\hline

\textbf{Memory interface} &
Demand training and cache fill interface. &
Throttling mechanisms tied to memory pressure. \\
\hline

\end{tabular}
\vspace{-.2in}
\end{table*}

\mm{Table~\ref{tab:integration_icp_vs_dvr_box} compares ICP and DVR in terms of integration with CPU core and memory hierarchy. Overall, ICP achieves substantially lower complexity, whereas DVR requires tight coupling with the CPU pipeline and thread support.}

\mm{\textbf{Thread requirement.} ICP does not require any additional hardware thread contexts. In contrast, DVR relies on a dedicated subthread to execute the decoupled runahead/discovery stream, introducing extra control for thread management.}

\mm{\textbf{Decode/execute modifications.} ICP incurs no changes to the core’s decode and execute stages. DVR, however, requires a mode-aware decode path (Discovery vs.\ DVR execution) and execution support for DVR-generated vectorized operations.}

\mm{\textbf{Commit requirements.} ICP only needs committed instruction information with loose timing. This enables a clean decoupling via an asynchronous buffer between the commit stage and ICP. DVR instead must ensure that DVR-generated operations do not update architectural state and must additionally provide termination/recovery logic when DVR stops.}

\mm{\textbf{Memory interface.} ICP uses a standard cache prefetcher interface. DVR, however, introduces additional throttling mechanisms to detect memory pressure and regulate the potentially high bandwidth demand generated by vectorized runahead.}

\subsection{Analysis of ICP Characteristics}

In this section, we examine two key characteristics of ICP: the \textbf{instruction correlation learning rate}, which reflects the time required to detect instruction correlations, and \textbf{the length of the derived dependency paths}, which represents the total number of instructions needed to compute the path from $\textit{PC}_\text{pre}$ to $\textit{PC}_\text{suc}$. A higher learning rate allows ICP to identify instruction correlations and initiate prefetching more quickly. Conversely, a shorter dependency path length enables ICP to compute the address of $\textit{PC}_\text{suc}$ more efficiently.

\subsubsection{Instruction Correlation Learning Rate}

\begin{figure}[t]
\centering
\includegraphics[width=.99\linewidth]{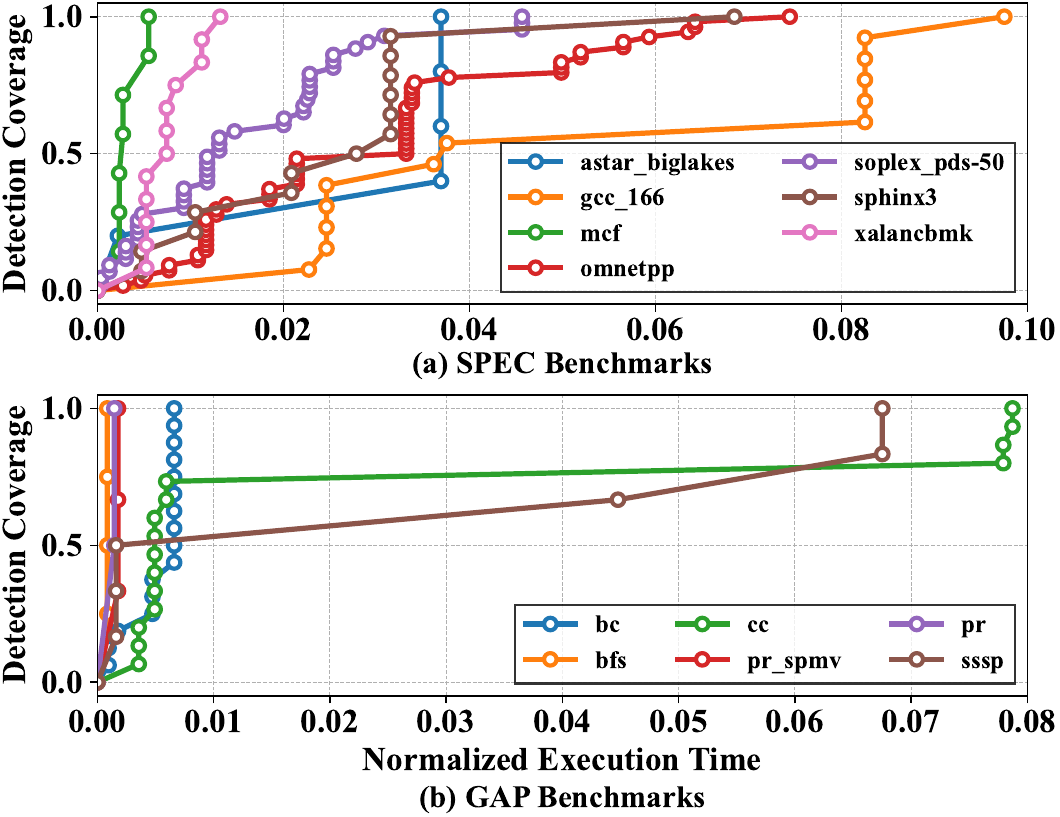}
\vspace{-.25in}
\caption{The learning rate of ICP. ICP completes the identification of all instruction correlations within less than 10\% of total program execution time.}
\vspace{-.15in}
\label{fig:learn_rate}
\end{figure}


Figure~\ref{fig:learn_rate} shows the detection coverage, defined as the ratio of covered correlations to the total number of identified correlations at the end of the programs, plotted against the program execution time normalized to the total execution time for ICP. The results demonstrate that ICP completes the identification of all instruction correlations within less than 10\% of total program execution time across both SPEC and GAP. Moreover, most correlations in SPEC are detected within the first 4\% of execution, while the majority in GAP are identified in under 1\%. By comparing the SPEC and GAP results, we find that the learning time increases with the complexity of memory access patterns. This is because instruction correlations may vary across different execution phases. ICP can only observe and record a correlation when the corresponding execution phase is encountered for the first time.

\subsubsection{Dependency Path Length}
\label{subsubsec:dpl}

\begin{figure}[t]
\centering
\includegraphics[width=.99\linewidth]{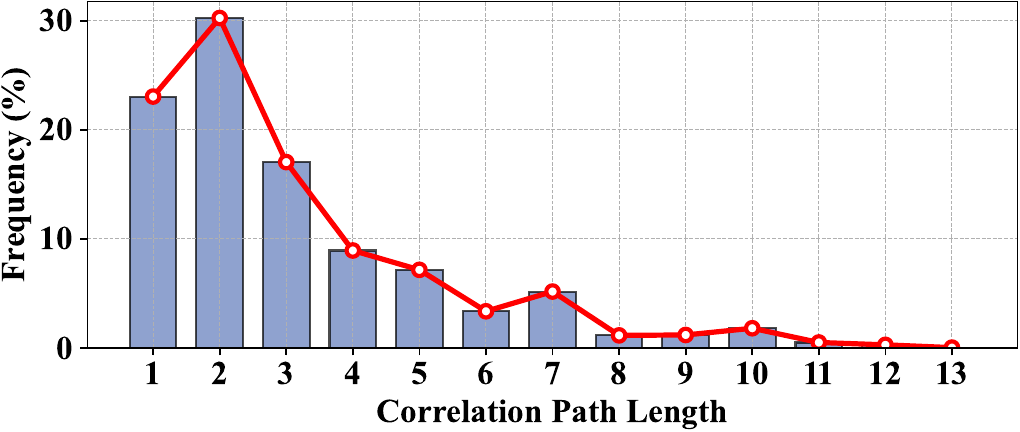}
\vspace{-.25in}
\caption{The dependency path length distribution. ICP completes the majority of its prefetching process within only a few cycles.}
\vspace{-.25in}
\label{fig:dep_dis}
\end{figure}

Figure~\ref{fig:dep_dis} shows the distribution of dependency path lengths across the SPEC and GAP benchmarks. The results reveal that most dependency paths are short: over 70\% of the identified paths have a length of no more than three instructions. This indicates that the majority of the speculation process initiated by ICP (i.e., the time from receiving cache responses from $\textit{PC}_\text{pre}$ to issuing prefetch requests for $\textit{PC}_\text{suc}$) requires the execution of fewer than 3 instructions. Additionally, the longest dependency path identified by ICP consists of only 13 instructions. Since ICP handles only basic arithmetic and logical operations (Section~\ref{subsec:pf_with_pcc}) and excludes all others, the speculative execution process typically takes between 1 to 3 cycles, with a maximum of 13 cycles. Given that cache prefetchers operate off the core’s critical execution path and typical memory access latencies span hundreds of cycles, this additional latency is negligible for prefetching timeliness and well within acceptable bounds.

\subsubsection{Correlation Table Size}
\label{subsubsec:corre_table_size}

\begin{figure}[t]
\centering
\includegraphics[width=.99\linewidth]{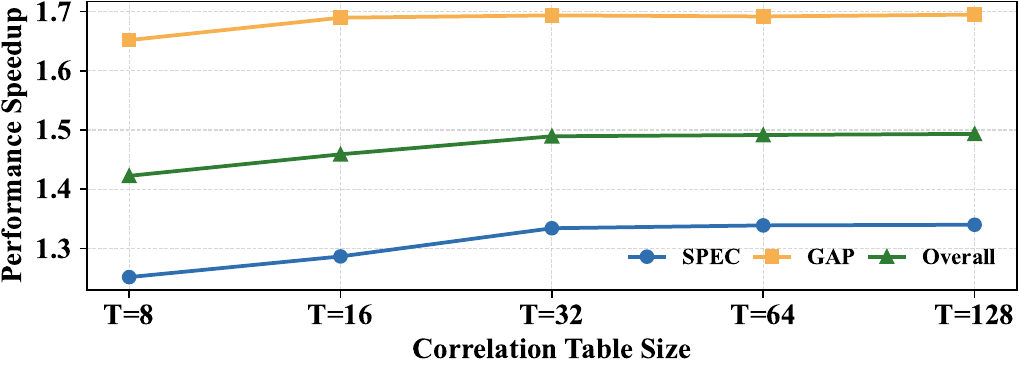}
\vspace{-.25in}
\caption{\mm{Performance impact of Correlation Table size.}}
\vspace{-.1 in}
\label{fig:corre_sen}
\end{figure}

\mm{Figure~\ref{fig:corre_sen} evaluates the performance impact of Correlation Table size. We vary the number of entries $T$ from 8 to 128. The results show that performance improves as the table size increases from $T=8$ to $T=32$, as a larger table can store more instruction-level correlations discovered during execution. However, increasing the table size further to $T=128$ yields only marginal improvement, demonstrating clear diminishing returns. The results highlight that the number of active instruction correlations within a program phase is relatively small, and a lighweight Correlation Table is sufficient to store them.}


\subsection{Ablation Study}
\label{subsec:ablation}

\begin{figure}[t]
\centering
\includegraphics[width=\linewidth]{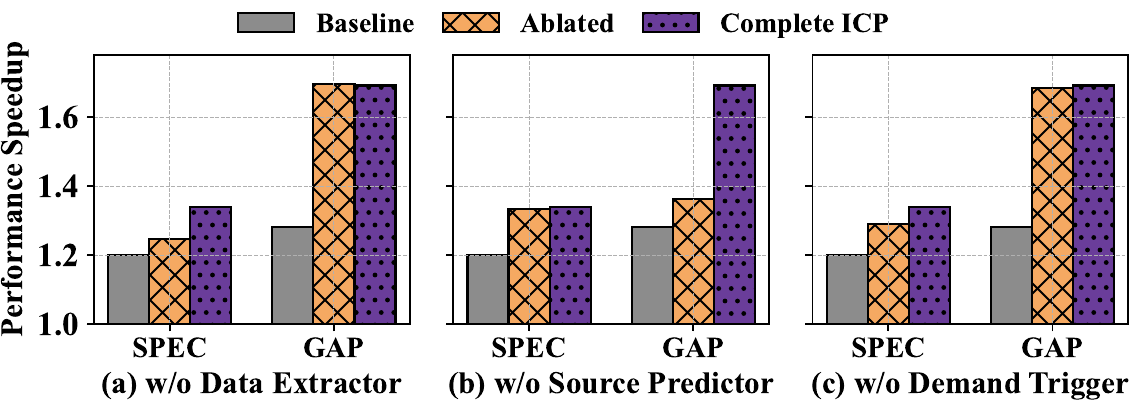}
\vspace{-.25in}
\caption{\mm{Ablation study of ICP.}}
\vspace{-.2in}
\label{fig:ablation}
\end{figure}

\mm{\textbf{Impact of Data Extractor.} Figure~\ref{fig:ablation}(a) shows performance on SPEC degrades noticeably when Data Extractor is removed. Data Extractor extracts the accessed value from prefetched cache lines by predicting the offset. Without it, ICP must consider all possible offsets within the cache line to extract values. This behavior significantly increases the number of unnecessary prefetches and leads to severe cache pollution. In contrast, $PC_{\text{pre}}$ in GAP often accesses multiple elements within a line with a fixed stride. Consequently, even when all values in the line are extracted, many of them remain useful.}

\mm{\textbf{Impact of Source Predictor.} Figure~\ref{fig:ablation}(b) shows Source Predictor has a particular impact on GAP, whose indirect patterns often involve dependency chains where intermediate operations consume two operands, with one operand being a base array value that remains stable but lies outside the chain. Without Source Predictor, ICP cannot supply these external operands during speculation.}

\textbf{Impact of Demand-Trigger Execution.}
Figure~\ref{fig:ablation}(c) shows that demand-triggered execution plays a secondary role compared to prefetch-triggered execution. The prefetch-triggered execution contributes about 9.02\% performance gain over the baseline for SPEC, whereas demand-triggered execution alone contributes about 4.92\% for SPEC. \mmi{The performance benefits enabled by demand-triggered execution stem from two main reasons.
First, although the dependency chain from $\textit{PC}_{\text{pre}}$ to $\textit{PC}_{\text{suc}}$ is short (Figure~\ref{fig:dep_dis}), the two instructions can be separated by a large number of control-flow dependent instructions, which may stall issue/execute due to resource contention (e.g., back-end pressure) and significantly extend the effective time gap between $\textit{PC}_{\text{pre}}$ and $\textit{PC}_{\text{suc}}$. In \texttt{gcc}, we observe cases where $\textit{PC}_{\text{suc}}$ executes only after several thousand cycles following $\textit{PC}_{\text{pre}}$.
Second, ICP executes the dependency chain upon receiving the cache-line response of $\textit{PC}_{\text{pre}}$, which is still earlier than when the core can execute the $\textit{PC}_{\text{suc}}$. When the core is blocked from issuing subsequent instructions due to resource contention, demand-triggered execution helps improve the prefetching timeliness.}

\subsection{Sensitivity Study for Temporal Prefetcher}
\label{subsec:sen_tp}



\begin{figure}[t!]
\centering
\includegraphics[width=.99\linewidth]{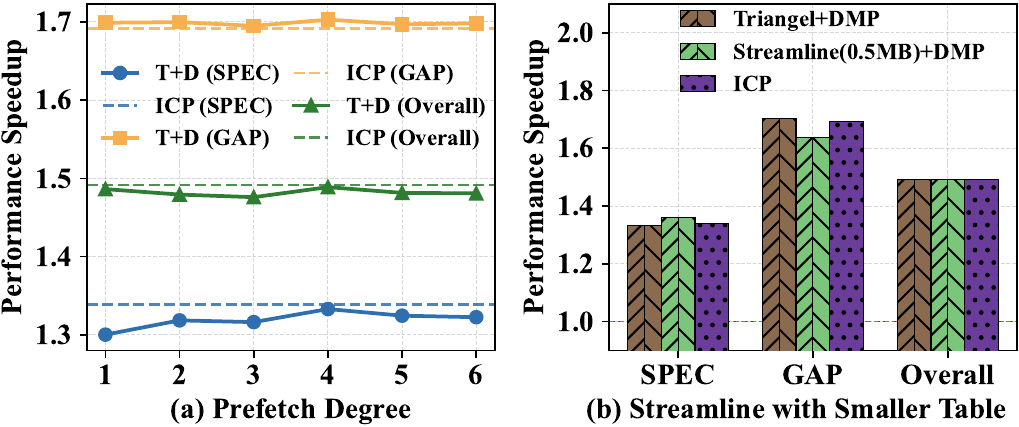}
\vspace{-.25in}
\caption{\mm{Sensitivity study for temporal prefetcher degree and table size.}}
\vspace{-.2in}
\label{fig:tp_sen}
\end{figure}

\mm{\textbf{Sensitivity to prefetch degree.} Figure~\ref{fig:tp_sen}(a) studies the sensitivity of DMP+Triangel to the prefetch degree of Triangel. We vary the degree from 1 to 6. The results show that performance remains relatively stable across different degrees, indicating that the hybrid design is not highly sensitive to this parameter. Among the evaluated settings, a degree of 4 achieves the best overall performance, and we use this configuration in our experiments.}

\mm{\textbf{Smaller metadata table.} Figure \ref{fig:tp_sen}(b) considers the state-of-the-art temporal prefetcher Streamline with a 0.5MB metadata table. Results show that this setup achieves better performance on SPEC, due to Streamline's efficient metadata compression, while it degrades on GAP. The degradation also stems from the compressed metadata, which binds multiple consecutive metadata targets with a single lookup address. GAP exhibits few memory-address repetition, making generating multiple prefetches within one lookup less reliable and issue more useless prefetches. Moreover, even when the metadata storage is reduced to 0.5~MB, the overhead remains several orders of magnitude larger than that of ICP.}

\subsection{Storage Overhead}
\label{subsec:storage}

\begin{table}[!h]
\centering
\vspace{-.25in}
\caption{Storage overhead of ICP.}
\vspace{-.1in}
\label{tab:storage}
\resizebox{0.99\columnwidth}{!}{%
\begin{tabular}{|c|c|c|}
\hline
\textbf{Structure} & \textbf{Entries} & \textbf{Storage} \\
\hline
\hline 
PC Selector\&Classifier & $8\times3.875B$ + $64 \times 1B$ & 95B 
\\
PC Correlation Detector & $128 \times 3.75B$ + $32 \times 1.75B$ & 536B 
\\
Correlation Table & $64 \times 7B$ & 448B 
\\
Data Extractor & $32 \times 392B$ & 150B 
\\
Source Predictor & $8 \times 9.25B$ & 74B
\\
Lightweight Calculator & N/A & 24B
\\
\mm{MSHR Extension} & N/A & 160B
\\
\mm{Commit FIFO Buffer} & $8 \times 16B$ & 128B
\\
\mmi{ROB Extension} & $288 \times 2B$ & 576B
\\
\hline
Total & N/A & 2.1KB
\\ \hline
\end{tabular}
}
\end{table}

\mmi{Table~\ref{tab:storage} shows ICP's storage overhead. Considering not all ROB implementations explicitly retain source register, we conservatively include the worst-case cost of augmenting the ROB with source register fields. Assuming a 288-entry ROB and an 8-bit register tag, storing two source register tags per entry incurs an additional $288 \times 2 \times 8=4{,}608$ bits ($\approx 576$B) of storage overhead. Two source register fields are sufficient because ICP only needs to model the basic operations supported by the Lightweight Calculator. These operations are at most binary and therefore require no more than two source operands. ICP's SRAM overhead sums to 2.1~KB ($17{,}200$ bits) across all structures.}

\mmi{Since 6T SRAM bitcell stores one bit using six transistors, this corresponds to $17{,}200\times 6 = 103{,}200$ bitcell transistors before accounting for any periphery. To express this in a technology-independent metric, we adopt the common gate-equivalent convention that uses a 2-input NAND (NAND2)~\cite{wikipedia_gate_equivalent} as the reference unit. As a result, the bitcells alone map to $103{,}200/4 \approx 25.8$k NAND2 gates. For small SRAM instances, decoders, wordline drivers, and sense amplifiers contribute non-trivial overhead. Therefore, we conservatively scale the bitcell-only estimate by $1.3$--$1.8\times$ to account for SRAM periphery, yielding $\sim$33.5k--46.4k NAND2.
Finally, adding the remaining non-SRAM logic (e.g., set-associative tag-compare/mux/control and the Lightweight Calculator datapath) results in an overall ICP budget of $\sim$40k--53k NAND2, which corresponds to a 6T-SRAM storage-equivalent capacity of
$(40\text{k--}53\text{k})\times 4/6/8 \approx 3.3$--4.3~KB.}


\subsection{Energy Overhead}
\label{subsec:energy}

We evaluate ICP's energy overhead on memory hierarchy level. Specifically, we use CACTI~\cite{muralimanohar2009cacti} to model the energy consumption of the on-chip memory hierarchy under $\qty{22}{nm}$. For DRAM accesses, we follow the methodology of Triangel~\cite{ainsworth2024triangel}, assuming the energy cost of a DRAM access is 25$\times$ that of an LLC access. \mm{The results show that ICP increases energy consumption by 2.4\%, 3.7\%, 1.8\%, and 7.3\% compared to Triangel, Tyche, DMP, and VR, respectively. In contrast, ICP reduces energy consumption by 6.0\% and 4.9\% compared to DVR and the DMP+Triangel, respectively.}

\mm{We further evaluate energy overhead introduced by ICP’s internal components. The dominant dynamic energy sources within ICP come from (1) accesses to its tables (e.g., Correlation Table) and (2) activations of Lightweight Calculator. To model the per-activation energy of Lightweight Calculator, we follow widely used energy reference points reported by Horowitz~\cite{horowitz20141}, which indicate that a simple integer add operation costs 0.1\,pJ. Accordingly, we model one Lightweight Calculator operation as 0.1\,pJ per activation.
For table accesses, \cite{horowitz20141} also reports that SRAM/cache accesses are at the pJ-level (e.g., $\sim$10\,pJ for a 64-bit access to an 8\,KB structure). Since ICP’s tables are much smaller than KB-scale caches, we bound one table access as 1\,pJ. Using measured event counts, we find that internal energy overhead of ICP is negligible, amounting to $0.24\%$ of the memory-hierarchy energy, and therefore does not materially affect overall energy conclusions.}

\section{Conclusion}
\label{sec:conclusion}

This paper presented ICP, a new hardware prefetching technique that exploits instruction-level correlations to address irregular memory access patterns. Unlike temporal prefetchers that rely on address recurrence, ICP learns stable correlations between instructions, enabling it to prefetch irregular accesses even when memory addresses never repeat. ICP outperforms state-of-the-art temporal prefetchers, indirect prefetchers, and runahead-based solutions, over three orders of magnitude less metadata than temporal prefetchers.
\section*{Acknowledgements}
We sincerely thank the anonymous reviewers of ISCA’26 for their insightful and constructive feedback. This work is supported by Hong Kong Research Grants Council (RGC) CRF-YCRG C6003-24Y, GRF 16216825, and T46-415/25-R. It was partially conducted by ACCESS – AI Chip Center for Emerging Smart Systems, supported by the InnoHK initiative of the Innovation and Technology Commission of the Hong Kong Special Administrative Region Government.


\bibliographystyle{IEEEtranS}
\bibliography{refs}


\end{document}